\def\year{2018}\relax
\documentclass[letterpaper]{article} 
\usepackage{aaai18}  
\usepackage{times}  
\usepackage{helvet}  
\usepackage{courier}  
\usepackage{url}  
\usepackage{graphicx}  

\usepackage{url}            
\usepackage{booktabs}       
\usepackage{amsfonts}       
\usepackage{nicefrac}       
\usepackage{microtype}      
\usepackage{subfigure} 
\usepackage{algorithm}
\usepackage{algorithmic}
\usepackage{bm}
\usepackage{amsmath}
\usepackage{amsfonts}
\usepackage{amssymb}

\usepackage{wrapfig}
\usepackage{color}

\begin{document}
%
\title{Video Generation From Text}
\author{
Yitong Li\\
Duke University \\
Durham, NC 27708 \\
\And
Martin Renqiang Min\\
NEC Labs America \\
Princeton, NJ 08540\\
\And 
Dinghan Shen\\
Duke University \\
Durham, NC 27708 \\
\And
David Carlson \\
Duke University \\
Durham, NC 27708 \\
\And
Lawrence Carin\\
Duke University \\
Durham, NC 27708 \\
}
\maketitle
\begin{abstract}

Generating videos from text has proven to be a significant challenge for existing generative models. We tackle this problem by training a conditional generative model to extract both static and dynamic information from text. This is manifested in a hybrid framework, employing a Variational Autoencoder (VAE) and a Generative Adversarial Network (GAN). The static features, called ``gist,'' are used to sketch text-conditioned background color and object layout structure. Dynamic features are considered by transforming input text into an image filter. To obtain a large amount of data for training the deep-learning model, we develop a method to automatically create a matched text-video corpus from publicly available online videos. Experimental results show that the proposed framework generates plausible and diverse videos, while accurately reflecting the input text information. It significantly outperforms baseline models that directly adapt text-to-image generation procedures to produce videos. Performance is evaluated both visually and by adapting the inception score used to evaluate image generation in GANs. 
    
 
\end{abstract}
\section{Introduction}\label{sec:intro}

Generating images from text is a well-studied topic, but generating videos based on text has yet to be explored as extensively. 
Previous work on the generative relationship between text and video has focused on producing text captioning from video~\cite{venugopalan2015sequence,donahue2015long,pan2016jointly,pu2016adaptive}. However, the inverse problem of producing videos from text has more degrees of freedom, and is a challenging problem for existing methods.
A key consideration in video generation is that both the broad picture and object motion must be determined by the text input.
Directly adapting text-to-image generation methods empirically results in videos in which the motion is not influenced by the text.


Video generation is related to video prediction.
In video prediction, the goal is to learn a nonlinear transfer function between given frames to predict subsequent frames~\cite{vondrick2017generating} -- this step is also required in video generation.
However, simply predicting future frames is not enough to generate a complete video.
Recent work on video generation has decomposed video into a static background, a mask and moving objects~\cite{vondrick2016generating,tulyakov2017mocogan}. 
Both of the cited works use a Generative Adversarial Network (GAN)~\cite{goodfellow2014generative}, which has shown encouraging results on sample fidelity and diversity.
%


However, in contrast with these previous works on video generation, here we conditionally generate videos based on side information, specially text captions.
Text-to-video generation requires both a good conditional scheme and a good video generator.
There are a number of existing models for text-to-image generation \cite{reed2016generative,mansimov2015generating};
unfortunately, simply replacing the image generator by a video generator provides poor performance (e.g., severe mode collapse), which we detail in our experiments.
These challenges reveal that even with a well-designed neural network model, directly generating video from text is difficult.

In order to solve this problem, we breakdown the generation task into two components. First, a conditional VAE model is used to generate the ``gist'' of the video from the input text, where the gist is an image that gives the background color and object layout of the desired video.
The content and motion of the video is then generated by conditioning on both the gist and text input.
This generation procedure is designed to mimic how humans create art.
Specifically, artists often draw a broad draft and then fill the detailed information. 
In other words, the gist-generation step extracts static and ``universal'' features from the text, while the video generator extracts the dynamic and ``detailed'' information from the text. 

\begin{figure*}[t]
\centering
\includegraphics[width=0.65\textwidth]{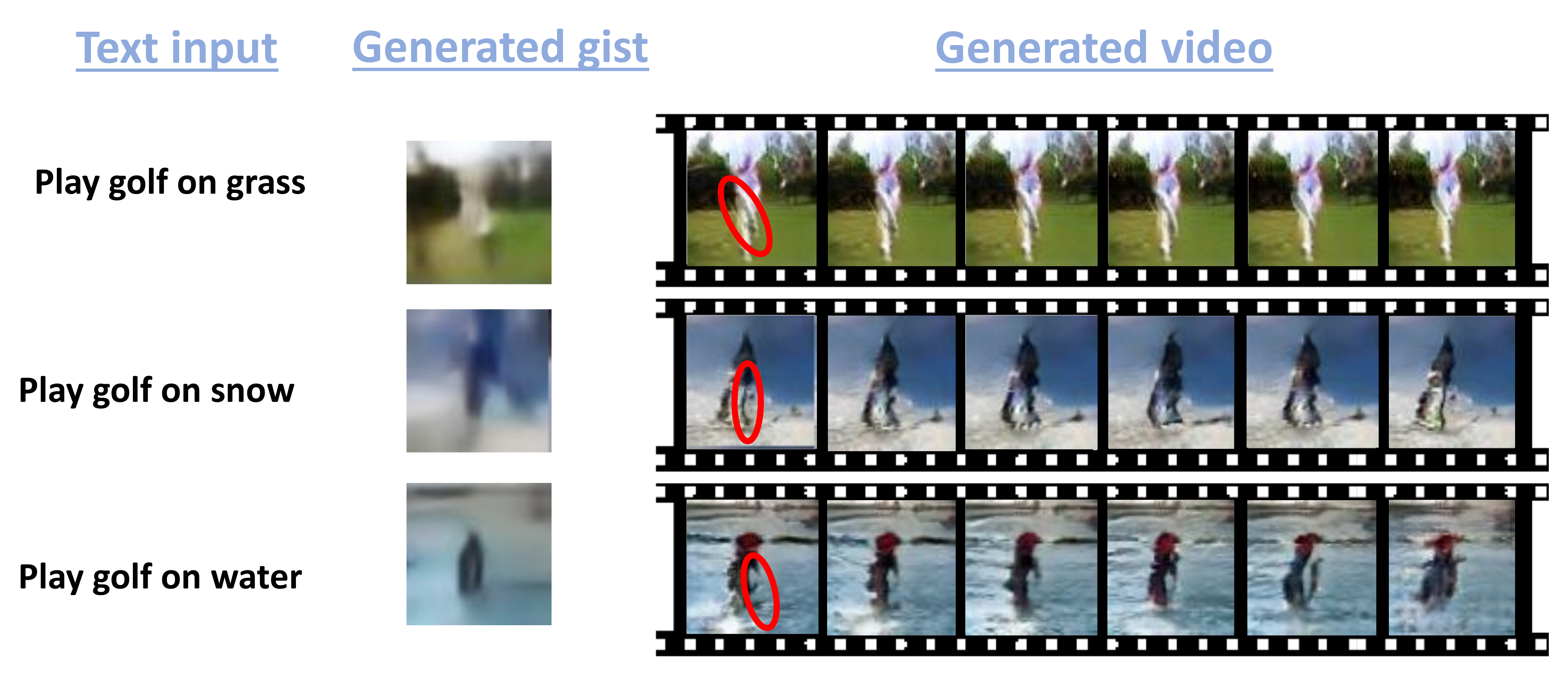}
\vspace{-3mm}
\caption{Samples of video generation from text. Universal background information (the gist) is produced based on the text.  The text-to-filter step generates the action (e.g., ``play golf''). The red circle shows the center of motion in the generated video.}
\label{fig:sample_1}
\end{figure*}

One approach to combining the text and gist information is to simply concatenate the feature vectors from the encoded text and the gist, as was previously used in image generation~\cite{yan2016attribute2image}. This method unfortunately struggles to balance the relative strength of each feature set, due to their vastly different dimensionality.
%
Instead, this work computes a set of image filter kernels based on the input text and applies the generated filter on the gist picture to get an encoded text-gist feature vector.
This combined vector better models the interaction between the text and the gist than simple concatenation, and is similar to the method used in~\cite{de2016dynamic} for video prediction and image-style transformation.
As we demonstrate in the experiments, the text filter better captures the motion information and adds detailed content to the gist.

Our contributions are summarized as follows: 
($i$) By viewing the gist as an intermediate step, we propose an effective text-to-video generation framework. 
($ii$) We demonstrate that using input text to generate a filter better models dynamic features. 
($iii$) We propose a method to construct a training dataset based on YouTube\footnote{\url{www.youtube.com}} videos where
the video titles and descriptions are used as the accompanying text.
This allows abundant on-line video data to be used to construct robust and powerful video representations.


\section{Related Work}\label{sec:related_work}

\subsection{Video Prediction and Generation}\label{subsec:related_video_gen}
Video generation is intimately related to video prediction. 
Video prediction focuses on making object motion realistic in a stable background.
Recurrent Neural Networks (RNNs) and the widely used sequence-to-sequence model~\cite{sutskever2014sequence} have shown great promise in these applications~\cite{villegas2017decomposing,de2016dynamic,van2017transformation,kalchbrenner2016video}. 
A common thread among these works is that a convolutional neural network (CNN) encodes/decodes each frame and connects to a sequence-to-sequence model to predict the pixels of future frames. 
In addition,~\cite{liu2017video} proposed deep voxel-flow networks for video-frame interpolation. 
Human-pose features have also been used to reduce the complexity of the generation~\cite{villegas2017decomposing,chao2017forecasting}.


There is also significant work on video generation conditioned on a given image.
Specifically, \cite{vukotic2017one,chao2017forecasting,walker2016uncertain,chen2017video,xue2016visual} propose methods to generate videos based on static images.
In these works, it is important to distinguish potential moving objects from the given image.
In contrast to video prediction, these methods are useful for generating a variety of potential futures, based upon the current image.
\cite{xue2016visual} inspired our work by using a cross-convolutional layer, where the motion is modeled by a set of image-dependent filter kernels.
The input image is convolved with its image-dependent kernels to give predicted future frames. 
A similar approach was also used in~\cite{de2016dynamic}.


GAN frameworks have been proposed for video generation without the need for a priming image.
A first attempt in this direction was made by \cite{vondrick2016generating}, who generated videos by separating scene and dynamic content.  
Using the GAN framework, a video could be generated purely from randomly sampled noise. 
Recently,~\cite{tulyakov2017mocogan} incorporated RNN model for video generation into a GAN-based framework, where noise is input at each time step. 
This model can construct a video simply by pushing random noise into an RNN model.

\begin{figure*}[ht]
\centering
\includegraphics[width=1.0\textwidth]{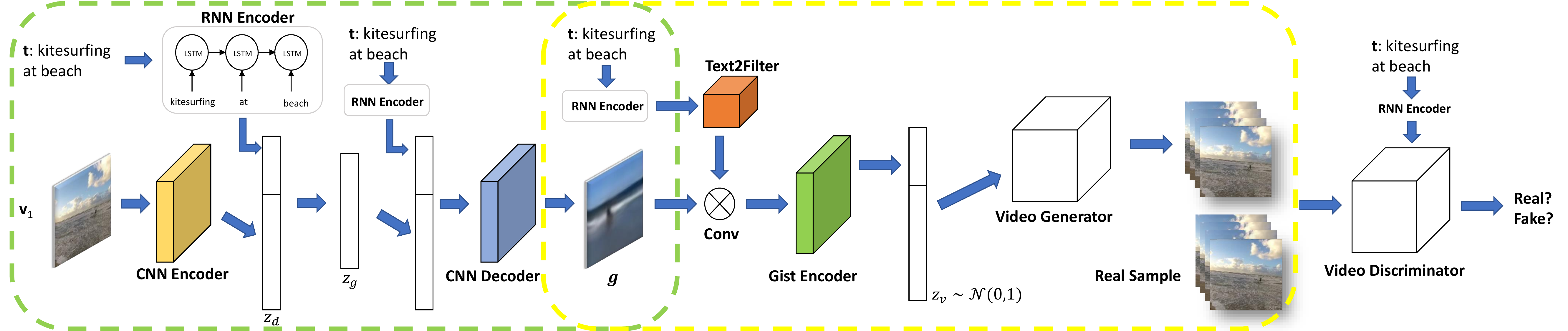}
\vspace{-3mm}
\caption{Framework of the proposed text-to-video generation method. The gist generator is in the green box. The encoded text is concatenated with the encoded frame to form the joint hidden representation $\bm z_d$, which further transformed into $\bm z_g$. The video generator is in the yellow box. The text description is transformed into a filter kernel (Text2Filter) and applied to the gist. The generation uses the features $\bm z_g$ with injected random noise. Following this point, the flow chart forms a standard GAN framework with a final discriminator to judge whether a video and text pair is real or synthetic. After training, the CNN image encoder is ignored. \vspace{-2mm}}

\label{fig:framework}
\end{figure*} 

\subsection{Conditional Generative Networks}\label{subsec:conditional_gan}
Two of the most popular deep generative models are the Variational Autoencoder (VAE)~\cite{kingma2013auto} and the Generative Adversarial Network (GAN)~\cite{goodfellow2014generative}. A VAE is learned by maximizing the variational lower bound of the observation while encouraging the approximate (variational) posterior distribution of the hidden latent variables to be close to the prior distribution. The GAN framework relies on a minimax game between a ``generator'' and a ``discriminator.'' The generator synthesizes data whereas the discriminator seeks to distinguish between real and generated data. In multi-modal situations, GAN empirically shows advantages over the VAE framework~\cite{goodfellow2014generative}.

In order to build relationships between text and videos, it is necessary to build conditionally generative models, which has received significant recent attention.
In particular, ~\cite{mirza2014conditional} proposed a conditional GAN model for text-to-image generation. The conditional information was given to both the generator and discriminator by concatenating a feature vector to the input and the generated image.
Conditional generative models have been extended in several directions. \cite{mansimov2015generating} generates images from captions with an RNN model using ``attention'' on the text.
~\cite{liu2016coupled,zhu2017unpaired} proposed conditional GAN models for either style or domain transfer learning. 
However, these methods focused on transfer from image to image.  Converting these methods to apply to text and image pairs is non-trivial.

The most similar work to ours is~\cite{reed2016generative}, which is the first successful attempt to generate images from text using a GAN model. 
In this work, pairs of data are constructed from the text features and a real or synthetic image. The discriminator tries to detect synthetic images or the mismatch between the text and the image. A direct adaptation of \cite{reed2016generative} unfortunately struggles to produce reasonable videos, as detailed in our experiments.
Text-to-video generation requires a stronger conditional generator than what is necessary for text-to-image generation. Video is a 4D tensor, where each frame is a 2D image with color information and spatiotemporal dependency. 
The increased dimensionality challenges the generator to extract both static and motion information from input text.
%

%

\section{Model Description}\label{sec:model}

We first introduce the components of our model, and then expand on each module in subsequent sections.
The overall structure of the proposed model is given in Figure~\ref{fig:framework}.
There are three model components: the conditional gist generator (green box), the video generator (yellow box), and the video discriminator.
The intermediate step of gist generation is developed using a conditional VAE (CVAE). Its structure is detailed in Section \ref{subsec:gist_gen}.
The video generation is based on the scene dynamic decomposition first proposed in~\cite{vondrick2016generating} with a GAN framework. The generation structure is detailed in Section \ref{subsec:video_gen}.
Because the proposed video generator is dependent on both the text and the gist, it is hard to incorporate all the information by a simple concatenation, as proposed by~\cite{reed2016generative}. 
Instead, this generation is dependent on a ``Text2Filter'' step described in Section \ref{subsec:text_gen}.
Finally, the video discriminator is used to train the model in an end-to-end fashion.


The data are a collection of $N$ videos and associated text descriptions, $\{ \bm V_i, \bm t_i \}$ for $i=1,\dots,N$.
Each video $\bm V_i\in\mathbb{R}^{T\times C \times H \times W}$ with frames $\bm V_i = \{ \bm v_{1i},\cdots, \bm v_{Ti} \}$, where $C$ reflects the number of color bands (typically $C=1$ or $C=3$), and $H$ and $W$ are the number of pixels in the height and width dimensions, respectively, for each video frame.
Note that all videos are cut to the same number of frames; this limitation can be avoided by using an RNN generator, but this is left for future work.
The text description $\bm t$ is given as a sequence of words (natural language).
The index $i$ is only included when necessary for clarity.

The text input was processed with a standard text encoder, which can be jointly trained with the model. Empirically, the chosen encoder is a minor contributer to model performance. Thus for simplicity, we directly adopt the skip-thought vector encoding model~\cite{kiros2015skip}.



\subsection{Gist Generator}\label{subsec:gist_gen}
In a short video clip, the background is usually static with only small motion changes. 
The gist generator uses a CVAE to produce the static background from the text (see example gists in Figure~\ref{fig:sample_1}). 
Training the CVAE requires pairs of text and images;
in practice, we have found that simply using the first frame of the video, $\bm v_1$, works well.

The CVAE is trained by maximizing the variational lower bound
\begin{eqnarray}\label{eq:gist_object}
\mathcal{L}_{gist} (\bm \theta_g, \bm \phi_g ; \bm v, \bm t) = \mathbb{E}_{q_{\phi_g} (\bm z_g | \bm v, \bm t)}\left[ \log p_{\theta_g}( \bm v | \bm z_g, \bm t ) \right] \cr
- \text{KL}\left( q_{\phi_g} (\bm z_g | \bm v, \bm t) || p ( \bm z_g) \right).
\end{eqnarray}
Following the original VAE construction~\cite{kingma2013auto}, the prior $p(\bm z_g)$ is set as an isotropic multivariate Gaussian distribution; $\theta_g$ and $\phi_g$ are parameters related to the decoder and encoder network, respectively. 
The subscript $g$ denotes gist. 
The encoder network $q_{\phi_g}(\bm z_g | \bm v, \bm t)$ has two sub-encoder networks $\eta(\cdot)$ and $\psi(\cdot)$. $\eta(\cdot)$ is applied to the video frame $\bm v$ and $\psi(\cdot)$ is applied to the text input $\bm t$.  
A linear-combination layer is used on top of the encoder, to combine the encoded video frame and text. Thus $\bm z_g \sim \mathcal{N} \left( \mu_{\phi_g}[\eta (\bm v);  \psi(\bm t)], \text{diag}\left( \sigma_{\phi_g}[\eta (\bm v);\psi(\bm  t)]\right)\right)$. 
The decoding network takes random noise $\bm z_g$ as an input.
The output of this CVAE network is called ``gist'', which is then one of the inputs to the video generator.

At test time, the encoding network on the video frame is ignored, and only the encoding network $\psi(\cdot)$ on the text is applied. This step ensures the model to get a sketch for the text-conditioned video.  In our experiments, we demonstrate that directly creating a plausible video with diversity from text is difficult. This intermediate generation step is critical.

\subsection{Video Generator}\label{subsec:video_gen}
The video is generated by three entangled neural networks, in a GAN framework, adopting the ideas of~\cite{vondrick2016generating}.
The GAN framework is trained by having a generator and a discriminator compete in a minimax game~\cite{goodfellow2014generative}.
The generator synthesizes fake samples to confuse the discriminator, while the discriminator aims to accurately distinguish synthetic and real samples.
This work utilizes the recently developed Wasserstein GAN formulation~\cite{arjovsky2017wasserstein}, given by
\begin{eqnarray}\label{eq:gan_loss}
\min_{\theta_G\in \Theta_G} \max_{\theta_D} \mathbb{E}_{\bm V \sim p(\bm V)} \left[ D(\bm V; \theta_D) \right] \cr 
- \mathbb{E}_{z_v \sim p(z_v)}\left[ D(G(\bm z_v; \theta_G); \theta_D) \right].
\end{eqnarray}
The function $D$ discriminates between real and synthetic video-text pairs, and the parameters $\theta_D$ are limited to maintain a maximum Lipschitz constant of the function. The generator $G$ generates synthetic samples from random noise that attempt to confuse the discriminator.  


As mentioned, conditional GANs have been previously used to construct images from text~\cite{reed2016generative}.
Because this work needs to condition on both the gist and text, it is unfortunately complicated to construct gist-text-video triplets in a similar manner.
Instead, first a motion filter is computed based on the text $\bm t$ and applied to the gist, further described in Section~\ref{subsec:text_gen}. 
This step forces the model to use the text information to generate plausible motion; simply concatenating the feature sets allows the text information to be given minimal importance on motion generation.
These feature maps are further used as input into a CNN encoder (the green cube in Figure~\ref{fig:framework}), as in \cite{isola2016image}. 
The output of the encoder is denoted by the text-gist vector $\bm g_t$, which jointly considers the gist and text information. 

To this point, there is no diversity induced for the motion in the text-gist vector, although some variation is introduced in the sampling of the gist based on the text information.
The diversity of the motion and the detailed information is primarily introduced by concatenating isometric Gaussian noise $\bm n_v$ with the text-gist vector, to form $\bm z_v = [\bm g_t; \bm n_v]$. The subscript $v$ is short for video.
The random-noise vector $\bm n_v$ gives motion diversity to the video and synthesizes detailed information.


We use the scene dynamic decomposition introduced in ~\cite{vondrick2016generating}. Given the vector $\bm z_v$, the output video from the generator is given by 
\begin{equation}\label{eq:video_generation}
G(\bm z_v) = \alpha(\bm z_v) \odot m(\bm z_v) + (1 - \alpha(\bm z_v)) \odot s(\bm z_v).
\end{equation}
The output of $\alpha(\bm z_v)$ is a 4D tensor with all elements constrained in $[0,1]$ and $\odot$ is element-wise multiplication. $\alpha(\cdot)$ and $m(\cdot)$ are both neural networks using 3D fully convolutional layers~\cite{long2015fully}. 
$\alpha(\cdot)$ is a mask matrix to separate the static scene from the motion. 
The output of $s(\bm z_v)$ is a static background picture repeated through time to match the video dimensionality, where the values in $s(\cdot)$ are from an independent neural network with 2D convolutional layers. 
Therefore, the text-gist vector $\bm g_t$ and the random noise combine to create further details on the gist (the scene) and dynamic parts of the video.


The discriminator function $D(\cdot)$ in \eqref{eq:gan_loss} is parameterized as a deep neural network with 3D convolutional layers; it has a total of five convolution and batch normalization layers. The encoded text is concatenated with the video feature on the top fully connected layer to form the conditional GAN framework.

\subsection{Text2Filter}\label{subsec:text_gen}

Simply concatenating the gist and text encoding empirically resulted in an overly reliant usage of either gist or text information.
Tuning the length and relative strength of the features is challenging in a complex framework.
Instead, a more-robust and effective way to utilize the text information is to construct the motion-generating filter weights based on the text information, which is denoted by Text2Filter.
This is shown as the orange cube in Figure~\ref{fig:framework}.

The filter is generated from the encoded text vector by a 3D convolutional layer of size
$F_c \times F_t \times kx \times ky \times kz$, where $F_t$ is the length of the encoded text vector. $F_c$ is number of output channels and $kx \times ky \times kz$ is filter kernel size. The 3D full-convolution~\cite{long2015fully} is applied to the text vector. In our experiments, $F_c = 64$. $kx=3$ in accordance with the RGB channels. $ky$ and $kz$ can be set arbitrarily, since they will become the kernel size of the gist after the 3D convolution. 
A deep network could also be adopted here if desired. 
Mathematically, the text filter is represented as
\begin{equation}\label{eq:text_filter}
f_g(\bm t) = \text{3Dconv} (\psi(\bm t)).
\end{equation}
Note that \textit{3Dconv} represents the 3D fully convolutional operation and $\psi(\cdot)$ is the text encoder.
The filter $f_g(\bm t)$ is directly applied on the gist to give the text-gist vector
\begin{equation}\label{eq:text-gist_vec}
\bm g_t = \text{Encoder} \left( \text{2Dconv} \left( \text{gist}, f_g(\bm t) \right) \right).
\end{equation}


\subsection{Objective Function, Training, and Testing}\label{subsec:model_train_test}
The overall objective function is manifested by the combination of $\mathcal{L}_{CVAE}$ and $\mathcal{L}_{GAN}$. 
Including an additional reconstruction loss $\mathcal{L}_{RECONS} = || \bm G - \hat{\bm V} ||_1$ empirically improves performance, where $\hat{\bm V}$ is the output of the video generator and $\bm G$ is $T$ repeats of $\bm g$ in time dimension. The final objective function is given by
\begin{equation}\label{eq:final_loss}
\mathcal{L} = \gamma_1 \mathcal{L}_{CVAE} + \gamma_2 \mathcal{L}_{GAN} + \gamma_3 \mathcal{L}_{RECONS} ,
\end{equation}
where $\gamma_1$, $\gamma_2$ and $\gamma_3$ are scalar weights for each loss term. In the experiments, $\gamma_1 = \gamma_2 = 1$ and $\gamma_3 = 0.1$, making the values of the three terms comparable empirically. The generator and discriminator are both updated once in each iteration. Adam~\cite{kingma2014adam} is used as an optimizer.

When generating new videos, the video encoder before $\bm z_g$ in Figure~\ref{fig:framework} is discarded, and the additive noise is drawn $\bm z_g \sim \mathcal{N}(\bm 0, \bm I)$. The text description and random noise are then used to generate a synthetic video.
%
%

\section{Dataset Creation}\label{sec:dataset}

Because there is no standard publicly available text-to-video generation dataset, we propose a way to download videos with matching text description. This method is similar in concept to the method in~\cite{ye2015eventnet} that was used to create a large-scale video-classification dataset.

Retrieving massive numbers of videos from YouTube is easy; however, automatic curation of this dataset is not as straightforward.
The data-collection process we have considered proceeds as follows. For each keyword, we first collected a set of videos together with their title, description, duration and tags from YouTube.
The dataset was then cleaned by outlier-removal techniques.
Specifically, the method of \cite{berg2010automatic} was used to get the $10$ most frequent tags for the set of video. 
The quality of the selected tags is further guaranteed by matching them to the words in existing categories in ImageNet~\cite{deng2009imagenet} and ActionBank~\cite{sadanand2012action}.
These two datasets help ensure that the selected tags have visually detectable objects and actions.
Only videos with at least three of the selected tags were included.
Other requirements include ($i$) the duration of the video should be within the range of $10$ to $400$ seconds, ($ii$) the title and description should be in English, and ($iii$) the title should have more than four meaningful words after removing numbers and stop words.


Clean videos from the Kinetics Human Action Video Dataset (Kinetics)~\cite{kay2017kinetics} are additionally used with the steps described above to further expand the dataset. The Kinetic dataset contains up to one thousand videos in each category, but the combined visual and text quality and consistency is mixed. 
For instance, some videos have non-English titles and others have bad video quality. 
In our experiments, we choose ten keywords as our selected categories: `biking in snow', `playing hockey', `jogging', `playing soccer ball', `playing football', `kite surfing', `playing golf', `swimming', `sailing' and `water skiing'. 
Note that the selected keywords are related to some categories in the Kinetic dataset. 
Most of the videos in the Kinetic dataset and the downloaded videos unfortunately have meaningless titles, such as a date indicating when the video was shot.
After screening these videos, we end up with about $400$ videos for each category. 
Using the YouTube8M~\cite{abu2016youtube} dataset for this process is also feasible, but the Kinetic dataset has cleaner videos than YouTube8M. 

\section{Experiments\footnote{All the generated samples in this section are uploaded as gif files in the Supplemental Material. We also include more generated videos for comparison. The Supplemental Material is publicly available at http://www.cs.toronto.edu/pub/cuty/Text2VideoSupp }}\label{sec:experiment}

\subsection{Video Preprocessing}\label{subsec:video_preprocess}
Current video-generation techniques only deal with smooth dynamic changes. 
A sudden change of shot or fast-changing background introduces complex non-linearities between frames, causing existing models to fail. 
Therefore, each video is cut and only qualified clips are used for the training~\cite{vondrick2016generating}.
The clips were qualified as follows. 
Each video uses a sampling rate of $2$5 frames per second. 
SIFT key points are extracted for each frame, and the RANSAC algorithm determines whether continuous frames have enough key-point overlap~\cite{lowe1999object}.
This step ensures smooth motions in the background and objects in the used videos.  
Each video clip is limited to $32$ frames, with $64 \times 64$ resolution. 
Pixel values are normalized to the range of $[-1,1]$, matching the use of the $tanh$ function in the network output layer.


\subsection{Models for Comparison}\label{subsec:baseline_models}
To demonstrate the effectiveness of our gist generation and conditional text filter, we compare the proposed method to several baseline models. The scene dynamic decomposition framework ~\cite{vondrick2016generating} is used in all the following baselines, which could be replaced with alternative frameworks. These baseline models are as follows:
\begin{figure}[t!]
    \centering
\subfigure[Baseline with only text encoder.]{\label{fig:baseline} \includegraphics[width=0.5\textwidth]{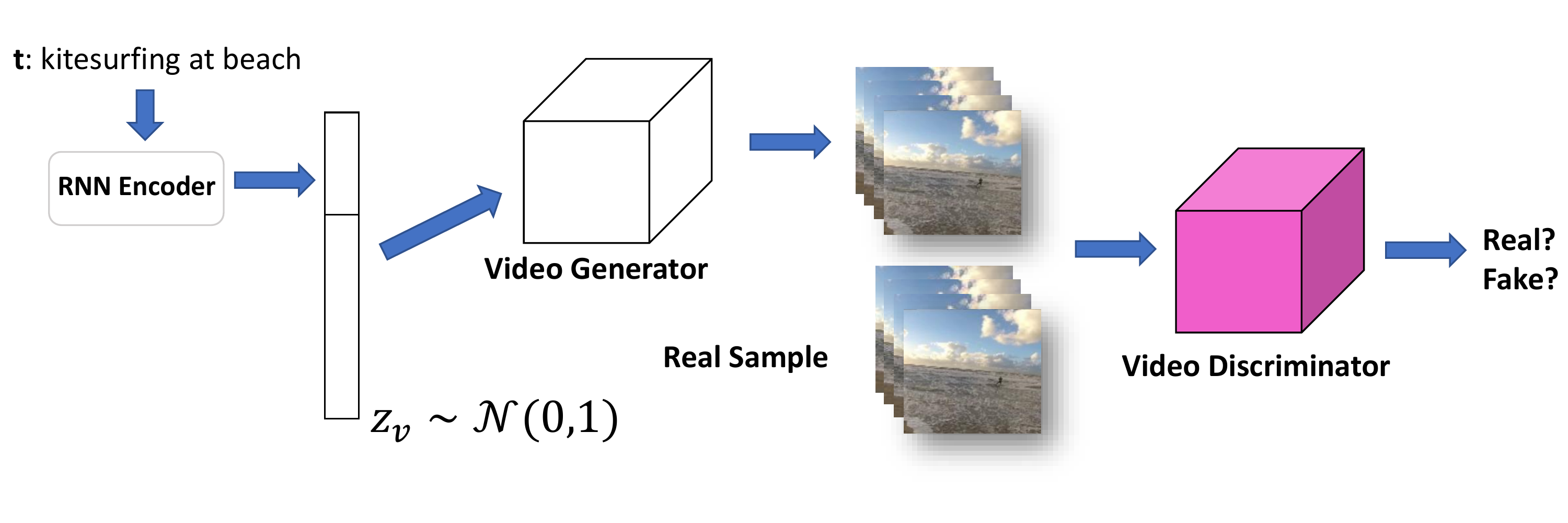} \vspace{-3mm}}
\subfigure[Baseline with pairing information.]{\label{fig:baseline_pair} \includegraphics[width=0.5\textwidth]{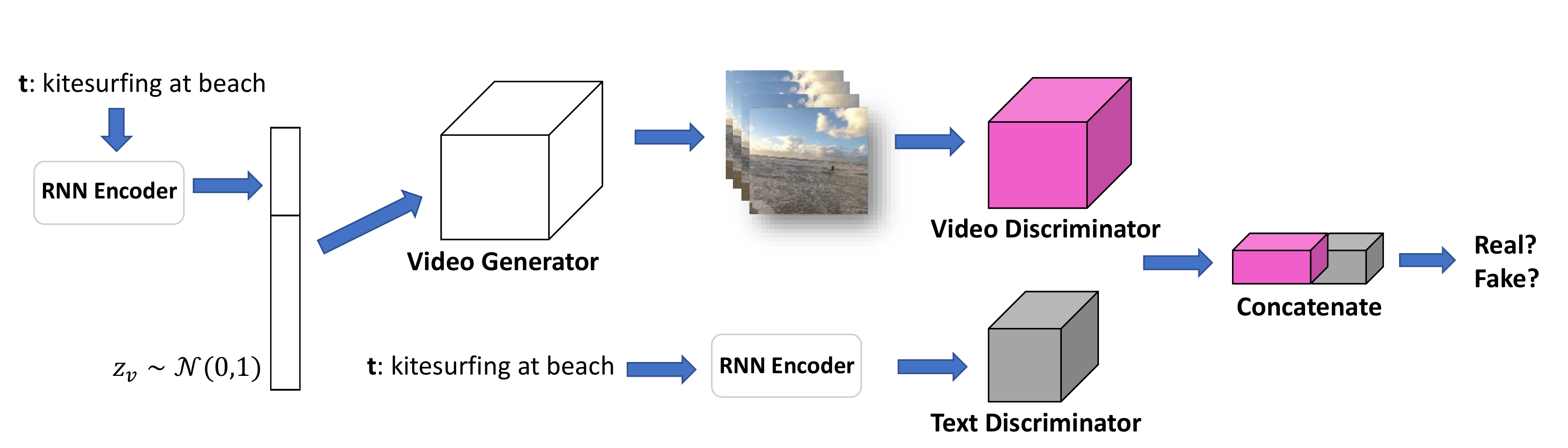}}
\caption{Two baselines adapted from previous work. Figure~\ref{fig:baseline} uses the conditional framework proposed by~\cite{vondrick2016generating}. The model was originally used for video prediction conditioned on a starting frame.  The starting frame in the model is replaced with text description. Figure~\ref{fig:baseline_pair} uses a discriminator performing on the concatenation of encoded video and text vectors. This is inspired by~\cite{reed2016generative}.}
\label{fig:baselines}
\end{figure}
\begin{figure*}[t!]
\centering
\includegraphics[width=0.9\textwidth]{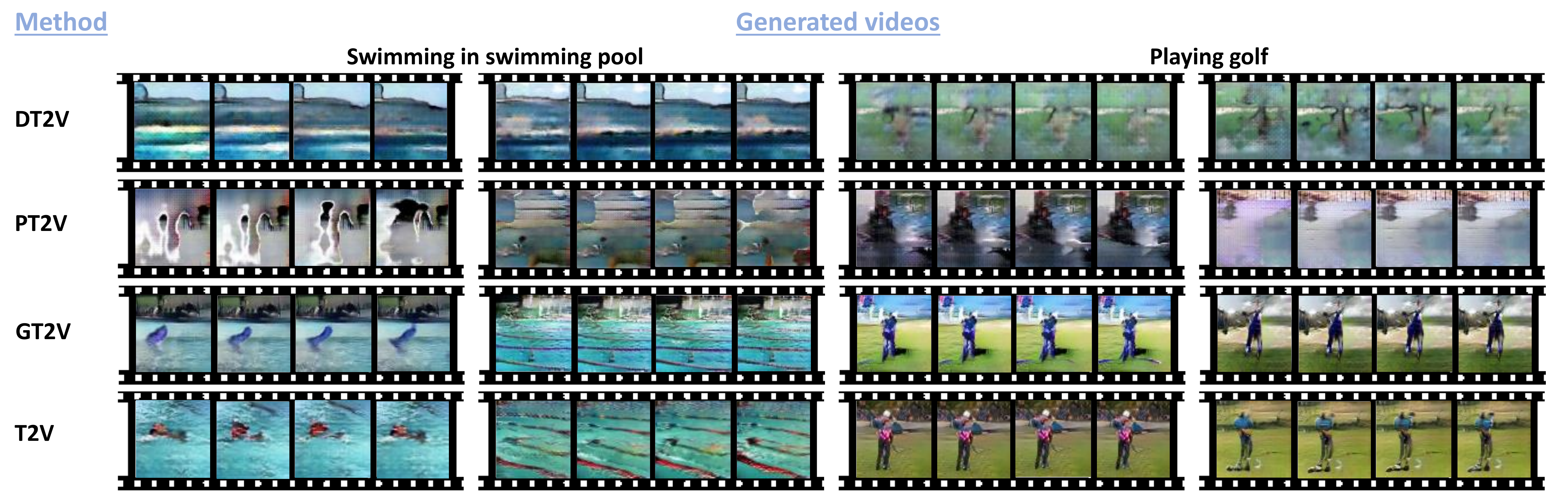}
\caption{Comparison of generated videos with different methods. The generated movie clips are given as supplemental files.}
\label{fig:sample_result}
\end{figure*} 

\begin{itemize}
\item \textbf{Direct text to video generation (DT2V)}: 
Concatenated encoded text $\psi(\bm t)$ and randomly sampled noise are fed into a video generator without the intermediate gist generation step.
This also includes a reconstruction loss $\mathcal{L}_{RECONS}$ in \eqref{eq:final_loss}. This is the method shown in Figure~\ref{fig:baseline}.
\item \textbf{Text-to-video generation with pair information (PT2V)}: DT2V is extended using the framework of~\cite{reed2016generative}. The discriminator judges whether the video and text pair are real, synthetic, or a mismatched pair. This is the method in Figure~\ref{fig:baseline_pair}. We use a linear concatenation for the video and text feature in the discriminator.
\item \textbf{Text-to-video generation with gist (GT2V)}: The proposed model, including only the conditional VAE for gist generation but {\em not} the conditional text filter (Text2Filter). 
\item \textbf{Video generation from text with gist and Text2Filter (T2V)} This is the complete proposed model in Section~\ref{sec:model} with both gist generation and Text2Filter components. 
\end{itemize}

Figure~\ref{fig:sample_result} presents samples generated by these four models, given text inputs ``swimming in the swimming pool'' and ``playing golf''.
The DT2V method fails to generate plausible videos, implying that the model in Figure~\ref{fig:baseline} does not have the ability to simultaneously represent both the static and motion features of the input. 
Using the ``pair trick'' from~\cite{reed2016generative} and ~\cite{isola2016image} does not drastically alter these results.
We hypothesize that, because the video is a 4D tensor while the text is a 1D vector, it renders balancing strength of each domain in the discriminator difficult.
By using gist generation, GT2V gives a correct background and object layout but is deficient in motion generation.
By concatenating the encoded gist vector, the encoded text vector, and the noise vector, the video generator of~\eqref{eq:video_generation} is hard to control.  Specifically, this method may completely ignore the encoded text feature when generating motion.
This is further explained in Section~\ref{subsec:qualitative_result}.
%


In comparison, the T2V model provides both background and motion features. The intermediate gist-generation step fixes the background style and structure, and the following Text2Filter step forces the synthesized motion to use the text information.
These results demonstrate the necessity of both the gist generator and the Text2Filter components in our model. In the following subsections, we intentionally generate videos that do not usually happen in real world. 
This is to address user concerns of simply replicating videos in the training set.

\subsection{Static Features}\label{subsec:result_static}
This section shows qualitative results of the gist generation, demonstrating that the gist reflects the static and background information from input text. 
\begin{figure}[htb]
\centering
\subfigure[Kitesurfing on the sea.]{\label{fig:sample_kiteonsea} \includegraphics[width=0.2\textwidth]{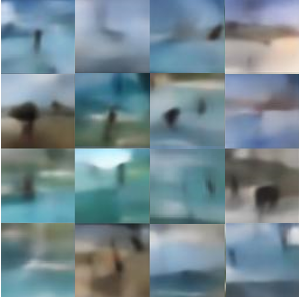}}
\subfigure[Kitesurfing on grass.]{\label{fig:sample_kiteongrass} \includegraphics[width=0.2\textwidth]{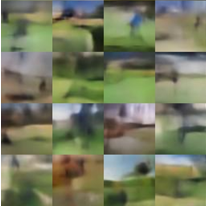}}
\subfigure[Swimming in swimming pool.]{\label{fig:sample_swiminpool} \includegraphics[width=0.2\textwidth]{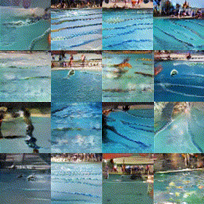}}
\subfigure[Swimming in snow.]{\label{fig:sample_swiminsnow} \includegraphics[width=0.2\textwidth]{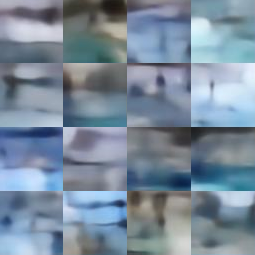}}
\caption{ Input text with same motion and different background information. The input text is given as the figure caption.}
\label{fig:gist_sample}
\end{figure}

\begin{figure}[t!]
\centering
\vspace{-1mm}
\includegraphics[width=0.44\textwidth]{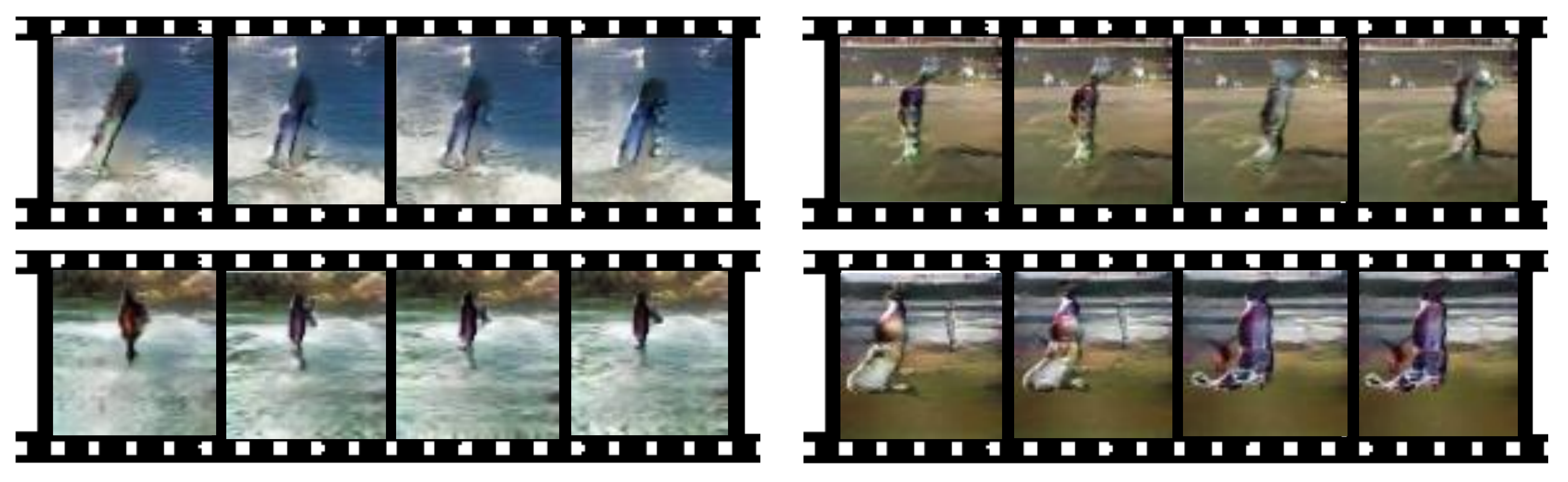}
\vspace{-2mm}
\caption{Left is ``kitesurfing on the sea''. Right is ``kitesurfing on grass''}
\label{fig:static_sample}
\end{figure} 

Figures \ref{fig:sample_kiteonsea} and \ref{fig:sample_kiteongrass} show sample gists of kite surfing at two different places. When generating videos with a grass field, the gist shows a green color. In contrast, when kite surfing on the sea, the background changes to a light blue. A black blurred shape appears in the gist in both cases, which is filled in with detail in the video generation. In Figure~\ref{fig:sample_swiminpool}, the lanes of a swimming pool are clearly visible. In contrast, the gist for swimming in snow gives a white background. Note that for two different motions at the same location, the gists are similar (results not shown due to space).

One of the limitations of our model is the capacity of motion generation. In Figure~\ref{fig:static_sample}, although the background color is correct, the kite-surfing motion on the grass is not consistent with reality. Additional samples can be found in Figure~\ref{fig:sample_1}.


\subsection{Motion Features}\label{subsec:result_dynamic}
\begin{figure}[tb]
\centering
\subfigure[Left is ``swimming at swimming pool''. Right is ``playing golf at swimming pool''.]{\label{fig:sample_motion1} \includegraphics[width=0.46\textwidth]{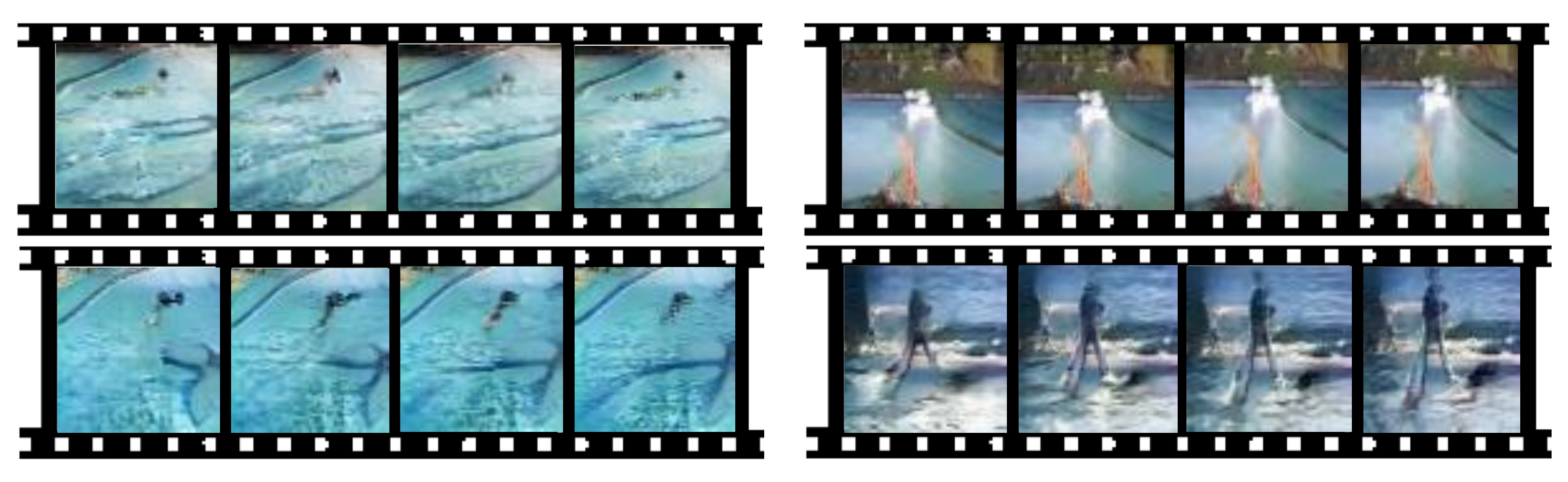}}
\subfigure[Left is ``sailing on the sea''. Right is ``running on the sea''.]{\label{fig:sample_motion2} \includegraphics[width=0.46\textwidth]{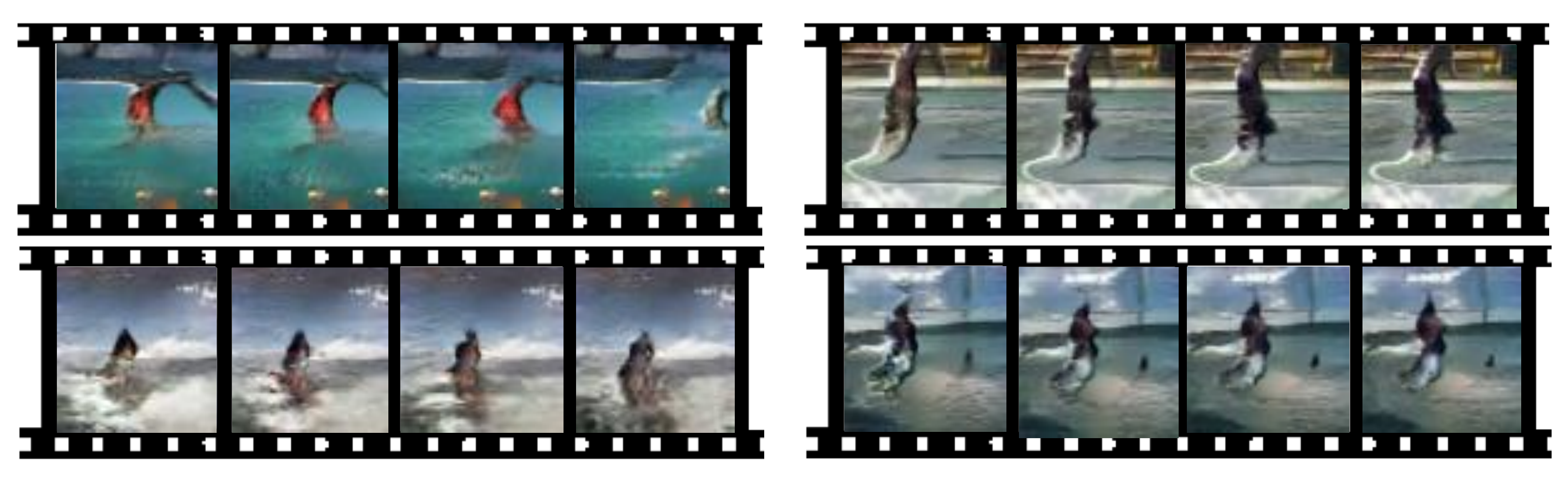}}
\vspace{-4mm}
\caption{Same textual motion for different locations. These texts inputs show generalization, as the right column do not exist in the training data.\vspace{-2mm}}
\label{fig:motion_sample}
\end{figure}

We further investigates motion-generation performance, which is shown by giving similar background and sampling the generated motion.
The samples are given in Figure~\ref{fig:motion_sample}.

This figure shows that a different motion can be successfully generated with similar backgrounds. However, the greatest limitation of the current CNN video generator is its difficulty in keeping the object shape while generating a reasonable motion. 
Moving to specific features such as human pose or skeleton generation could provide improvements to this issue ~\cite{chao2017forecasting,walker2017pose}.

\subsection{Quantitative Results}\label{subsec:qualitative_result}
Following the idea of inception score~\cite{salimans2016improved}, we first train a classifier on six categories: `kite surfing', `playing golf', `biking in snow', `sailing', `swimming' and `water skiing.'
Additional categories were excluded due to the low in-set accuracy of the classifier on those categories. 

A relatively simple video classifier is used, which is a five-layer neural network with 3D full convolutions~\cite{long2015fully} and ReLU nonlinearities.
The output of the network is converted to classification scores through a fully connected layer followed by a soft-max layer.
In the training process, the whole video dataset is split with ratios $7:1:2$ to create training, validation and test sets. 
The trained classifier was used on the $20\%$ left-out test data as well as the generated samples from the proposed and baseline models. The classification accuracy is given in Table~\ref{tab:acc_comparison}.

\begin{table}[t]
  \resizebox{1\hsize}{!}{
  \begin{tabular}{l|c|c|c|c|c}
  \hline
    & In-set & DT2V &PT2V &GT2V & T2V  \\
    \hline
    Accuracy & 0.781 & 0.101 & 0.134 &  0.192 & 0.426 \\
   \hline 
  \end{tabular}
  }
  \caption{Accuracy on different test sets. `In-set' means the test set of real videos. DT2V, PT2V, GT2V, and T2V (the full proposed model) are described in Section~\ref{subsec:baseline_models}.}
\label{tab:acc_comparison}
\end{table}

We observe clear mode collapse when using D2T and PT2V, explaining their poor performance.
Further, it appears that directly generating video from a GAN framework fails because the video generator is not powerful enough to account for both the static and motion features from text.
Using the gist generation in GT2V provides an improvement over the other baseline models. This demonstrates the usefulness of gist, which alleviates the burden of the video generator. 
Notably, the full proposed model (including Text2Filter) performs best on this metric by a significant margin, showing the necessity of both the gist generation and Text2Filter.


\begin{figure}[tb]
\centering
\vspace{-4mm}
\includegraphics[width=0.34\textwidth]{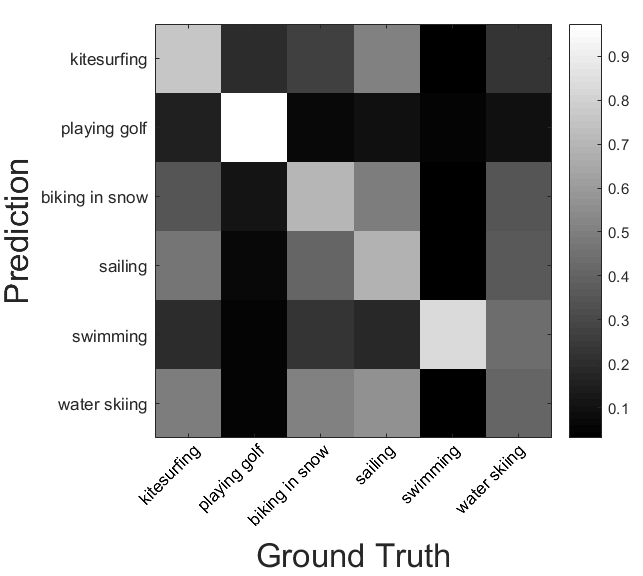}
\vspace{-4mm}
\caption{Classification confusion matrix on T2V generated samples.}
\label{fig:classification}
\end{figure} 
Figure~\ref{fig:classification} shows the confusion matrix when the classifier is applied to the generated videos of our full model. Generated videos of swimming and playing golf are easier to classify than other categories. In contrast, both `sailing' and `kite surfing' are on the sea. Thus it is difficult to distinguish between them. This demonstrates that the gist generation step distinguishes different background style successfully.

\section{Conclusion}\label{sec:conclusion}
This paper proposes a framework for generating video from text using a hybrid VAE-GAN framework. 
To the best of our knowledge, this work proposes the first successful framework for video generation from text. 
The intermediate gist-generation step greatly helps enforce the static background of video from input text. 
The proposed Text2Filter helps capture dynamic motion information from text. 
In the future, we plan to build a more powerful video generator by generating human pose or skeleton features, which will further improve the visual quality of generated human activity videos.

{\small
\bibliographystyle{aaai}
\bibliography{main}
}

\end{document}